\documentclass[fleqn,twoside,twocolumn,nofootinbib,showkeys]{revtex4} 
\usepackage[sec,nocpr]{ujp} 

\begin{document}
\title[STRUCTURAL INSTABILITY OF A BIOCHEMICAL PROCESS]
{STRUCTURAL INSTABILITY OF A BIOCHEMICAL PROCESS}%
\author{V. Grytsay}
\affiliation{\bitp}
\address{\bitpaddr}

\udk{577.3} \pacs{05.45.-a, 05.45.Pq,\\[-3pt] 05.65.+b} \razd{\secx}

\autorcol{V.\hspace*{0.7mm}Grytsay}

\begin{abstract}
By the example of a mathematical model of a biochemical process,
the structural instability of dynamical systems is studied by
calculating the full spectrum of Lyapunov indices with the use of
the generalized Benettin algorithm. For the reliability of the
results obtained, the higher Lyapunov index determined with the
orthogonalization of perturbation vectors by the Gram--Schmidt
method is compared with that determined with the overdetermination
of only the norm of a perturbation vector. Specific features of
these methods and the comparison of their efficiencies for a
multidimensional phase space are presented. A scenario of the
formation of strange attractors at a change of the dissipation
parameter is studied. The main regularities and the mechanism of
formation of a deterministic chaos due to the appearance of a fold
or a funnel, which leads to the uncertainty of the evolution of a
biosystem, are determined.
\end{abstract}

\maketitle

\section{Introduction}

    One of the main physical problems is the appearance of ordered structures
from the chaos in systems different by their nature due to the
self-organization. The synergy was follow from theoretical physics
[1]. The first model of synergy was the Turing model (``Morphogenesis
model'') [2]. The next was the Prigogine's ``Brusselator'', where
self-organization regimes were considered in an abstract
chemicothermal system [3]. Synergy allows one to find common physical rules
for the self-organization of opened nonlinear systems [4--7].

To a significant extent, the problem concerns the question of the
self-origination of life, the evolutionary development of the alive,
and the basic mechanisms of structural-functional regularities of
transformations in various biochemical systems. In the general case,
biochemical systems are described by ordinary nonlinear differential
equations of the form ${d \overline X \over dt} = \overline f
(\overline X ,\overline a )$, where $ \overline X = (X_1,...X_n)\in
\overline {R^n} $  is a vector of variables of states (phase
variables), and $ \overline a = (a_1,...,a_k) \in \overline {R^k})$
is the vector of parameters of the system. The results of numerical
solutions of the equations can be compared with experiments and would clarify
self-organization laws.

Works [8--18] considered the mathematical model of a bioreactor
transforming steroids [19] under flow conditions depending on a
change of the dissipation, the kinetic membrane potential of
cells, and the input flows of a substrate and oxygen. Various
scenarios of the transition from stationary modes to
self-oscillatory modes with different multiplicities were
presented, and the regions of the formation of strange attractors
were determined. It is worth noting an experiment that proved the existence of self-oscillations
in a population of {\it Arthrobacter globiformis} cells [20].

The studies were performed with the use of the method of phase
portraits. The determined regions with qualitatively identical
phase portraits and the points of bifurcation do not characterize
the dynamics of a biosystem sufficiently completely. The most
complete information about the stability of various modes is
contained in the full spectrum of Lyapunov indices. But since a
mathematical model of the given biochemical system contains a lot
of variables and parameters, the limitations on the solution of
such problems on a computer arise due to a small volume of the
work memory for the processing of a matrix of small
perturbations. In addition, any error made in the programming will
essentially influence the overdetermination of perturbation
vectors and their orthogonalization.

To attain the reliable results, we carried out the independent
calculations of both the higher Lyapunov index with the same
parameters, by using the Benettin algorithm with the
overdetermination of only the norm of perturbation vectors, and
the full spectrum of Lyapunov indices with the orthogonalization
of these vectors by the Gram--Schmidt method [21--23]. The higher
Lyapunov indices obtained were practically identical, which
confirms the correctness of the developed computer program.

The essence of the calculation of a higher Lyapunov index with the
overdetermination of only the norm of perturbation vectors consists in the
determination of the evolution of an arbitrarily small deviation from a studied
trajectory of the system
${\lambda = {1\over n\tau}\sum_{k=1}^n} \ln{\parallel \overline {u_k} \parallel
 \over \varepsilon}$.
After each step of calculations, it is necessary to overdetermine a deviation
so that its direction will remain the same, and the norm will be equal to the
input value $\varepsilon$, namely: ${ \overline {u_0k}} =
{\varepsilon \overline {u_k} \over \parallel \overline {u_k} \parallel}$.

The algorithm of calculations of the full spectrum of Lyapunov
indices consisted in the following. Taking some point on the
attractor $ \overline {X_0}$ as the initial one, we traced the
trajectory outgoing from it and the evolution of $N$ perturbation
vectors. In our case, $N = 10$ (the number of variables of the
system [18]). The initial equations of the system supplemented by 10
complexes of equations in variations were solved numerically. As the
initial perturbation vectors, we set the collection of vectors
$\overline {b^0_1}$, $ \overline {b^0_2}$,... $ \overline
{b^0_{10}}$ which are mutually orthogonal and normed by one. In
some time $T$, the trajectory arrives at a point $ \overline X_1$,
and the perturbation vectors become $ \overline {b^1_1}$, $
\overline {b^1_2}$,... $ \overline {b^1_{10}}$, Their
renormalization and orthogonalization by the Gram--Schmidt method are
performed by the following scheme:
\[
\overline {b^1_1} =
\frac{\overline {b_1}}{\parallel \overline {b_1} \parallel},
\]
\[
\overline {b^\prime_2} = \overline {b^0_2} -
(\overline {b^0_2},\overline {b^1_1}) \overline {b^1_1},\
\overline {b^1_2} =
\frac{\overline {b^\prime_2}}{\parallel \overline {b^\prime_2} \parallel},
\]
\[
\overline {b^\prime_3} = \overline {b^0_3} -
(\overline {b^0_3},\overline {b^1_1})
\overline {b^1_1} - (\overline {b^0_3},\overline {b^1_2}) \overline {b^1_2},\
\overline {b^1_3} =
\frac{\overline {b^\prime_3}}{\parallel \overline {b^\prime_3} \parallel},
\]
\[
\overline {b^\prime_4} = \overline {b^0_4} -
(\overline {b^0_4},\overline {b^1_1})
\overline {b^1_1} - (\overline {b^0_4},\overline {b^1_2}) \overline {b^1_2} -
(\overline {b^0_4},\overline {b^1_3}) \overline {b^1_3},\
\overline {b^1_4} =
\frac{\overline {b^\prime_4}}{\parallel \overline {b^\prime_4} \parallel},
\]
\[
...............................................................................
.......\]
\centerline{}

Then the calculations are continued, by starting from the point $\overline X_1$
and perturbation vectors
$\overline {b^1_1}$, $ \overline {b^1_2}$,... $ \overline {b^1_{10}}$.
After the next time interval $T$, a new collection of perturbation vectors
$\overline {b^2_1}$, $ \overline {b^2_2}$,... $ \overline {b^2_{10}}$ is formed
and undergoes again the orthogonalization and renormalization by the
above-indicated scheme. The described sequence of manipulations is repeated
a sufficiently large number of times, $M$. In this case in the course of
calculations, we evaluated the sums
\[
S_1 =\sum_{i=1}^M \ln \parallel b^{\prime i}_1 \parallel, \ S_2
=\sum_{i=1}^M \ln \parallel b^{\prime i}_2 \parallel,...,
\]
\[
S_{10} =\sum_{i=1}^M \ln \parallel b^{\prime i}_{10} \parallel,
\]
which involve the perturbation vectors prior to the renormalization, but after
the normalization.  The estimation of 10 Lyapunov indices was carried out
in the following way:
\[
\lambda_j = {S_j \over {MT}},\ i=1,2,...10.
\]

As the test calculations for the verification of a program, we reproduced
the well-known results for the finite-dimensional Lorentz system.

\section{Mathematical Model}

A mathematical model of the process under flow conditions in a bioreactor was
developed by the general scheme of metabolic processes in {\it Arthrobacter globiformis}
cells at a transformation of steroids [8-18].
\begin{equation}
\frac{dG}{dt} =  \frac{G_0}{N_3 + G + \gamma_2\Psi} - l_1
V(E_1)V(G) - \alpha_3G ,
\end{equation}
\begin{equation}
\frac{dP}{dt} =  l_1 V(E_1)V(G) - l_2V(E_2)V(N)V(P) -
\alpha_4P,
\end{equation}
\begin{equation}
\frac{dB}{dt} = l_2 V(E_2)V(N) V(P) - k_1V(\Psi)V(B) -
\alpha_5B,
\end{equation}
\[
\frac{dN}{dt} = - l_2V(E_2)V(P)V(N) - l_7V(Q)V(N) + \]
\begin{equation}
+ k_{16}V(B) \frac{\Psi}{K_{10} + \Psi} +
\frac{N_0}{N_4 + N} - \alpha_6N,
\end{equation}
\[
\frac{d{E_1}}{dt} = E_{10}\frac{G^2}{\beta_1 + G^2}\left(1 -
\frac{P + mN}{N_1 + P+mN}\right) - \]
\begin{equation}
- l_1V(E_1)V(G) + l_4V(e_1)V(Q) - a_1E_1,
\end{equation}
\begin{equation}
\frac{d{e_1}}{dt} = - l_4V(e_1)V(Q) + l_1V(E_1)V(G) -
\alpha_1e_1,
\end{equation}
\[
\frac{dQ}{dt} = 6lV(Q^0 + q^0 - Q)V(O_2)V^{(1)}(\Psi) - \]
\begin{equation}
- l_6V(e_1)V(Q) - l_7V(Q)V(N),
\end{equation}
\[
\frac{d{O_2}}{dt} = \frac{O_{20}}{N_5 + O_2} - lV(O_2)V(Q^0
+ q^0 - Q)V^{(1)}(\Psi) - \]
\begin{equation}
- \alpha_7O_2,
\end{equation}
\[
\frac{d{E_2}}{dt} = E_{20} \frac{P^2}{\beta_2 + P^2}
\frac{N}{\beta + N}(1- \frac{B}{N_2 + B} - \]
\begin{equation}
- l_{10}V(E_2)V(N)V(P) - \alpha_2E_2,
\end{equation}
\begin{equation}
\frac{d\Psi}{dt} = l_5V(E_1)V(G) + l_9V(N)V(Q) - \alpha\Psi.
\end{equation}

\noindent where: $V(X) =X/(1 + X)$;  $V^{(1)}(\Psi) =1/( 1 +
\Psi^2)$; $V(X)$ is a function involving the adsorption of an
enzyme in the region of a local bond; $V^{(1)}(\Psi)$ is a function
characterizing the influence of the kinetic membrane potential on
the respiratory chain.  In the modeling, it is convenient to use the
following dimensionless quantities [1--11] which are set as follows:
$l = l_1 = k_1 = 0.2;$ $l_2 = l_{10} = 0.27$; $l_5 = 0.6$; $l_4 =
l_6 = 0.5$; $l_7 = 1.2$; $l_9 = 2.4$; $k_2 = 1.5$; $E_{10} = 3$;
$\beta_1 = 2$; $N_1 = 0.03$; $m = 2.5$; $\alpha = 0.0033$; $a_1 =
0.007$; $\alpha_1= 0.0068$; $E_{20} = 1.2$; $\beta = 0.01$; $\beta_2
= 1$; $N_2 = 0.03$; $\alpha_2 = 0.02$; $G_0 = 0.019$; $N_3 = 2$;
$\gamma_2 = 0.2$; $\alpha_5 = 0.014$; $ \alpha_3 = \alpha_4 =
\alpha_6 = \alpha_7 = 0.001$; $O_{20} = 0.015$; $N_5 = 0.1$; $N_0 =
0.003$; $N_4 = 1$; $K_{10} = 0.7$.

Equations (1)--(9) describe a change in the concentrations of (1)
-- hydrocortisone $(G)$; (2) -- prednisolone $(P)$; (3) --
$20\beta$-oxyderivative of prednisolone $(B)$; (4) -- reduced form
of nicotinamideadeninedinucleotide $(N)$; (5) -- oxidized form of
3-ketosteroid-$\bigtriangleup$-dehydrogenase $(E_1)$; (6) --
reduced form of 3-ketosteroid-$\bigtriangleup$-dehydrogenase
$(e_1)$; (7) -- oxidized form of the respiratory chain $(Q)$; (8)
-- oxygen $(O_2)$; (9) -- $20\beta$-oxysteroid-dehydrogenase
$(E_2)$. Equation (10) describes a change in the kinetic membrane
potential $(\Psi)$.

The initial parameters of the system are as
follows: $G^0 = 0.17$; $P^0 = 0.844$; $B^0 = 0.439$; $N^0 = 1.789$;
$E^0_1 = 0.216$; $e^0_1 = 1.835$; $Q^0 = 2.219$; $O^0_2 = 0.309$;
$E^0_2 = 1.645$; $\Psi^0 = 0.300$.

The reduction of parameters of the system to dimensionless
quantities is given in works [8,9].  To solve this autonomous
system of nonlinear differential equations, we applied the
Runge--Kutta--Merson method. The accuracy of solutions was set to be
$10^{-12}$. To get the reliable results, namely in order that the
system, being in the initial transient state, approach the
asymptotic attractor mode, we took the duration of calculations to
be 100000. For this time interval, the trajectory ``sticks'' the
corresponding attractor.

\begin{figure*}
\includegraphics[width=17.5cm]{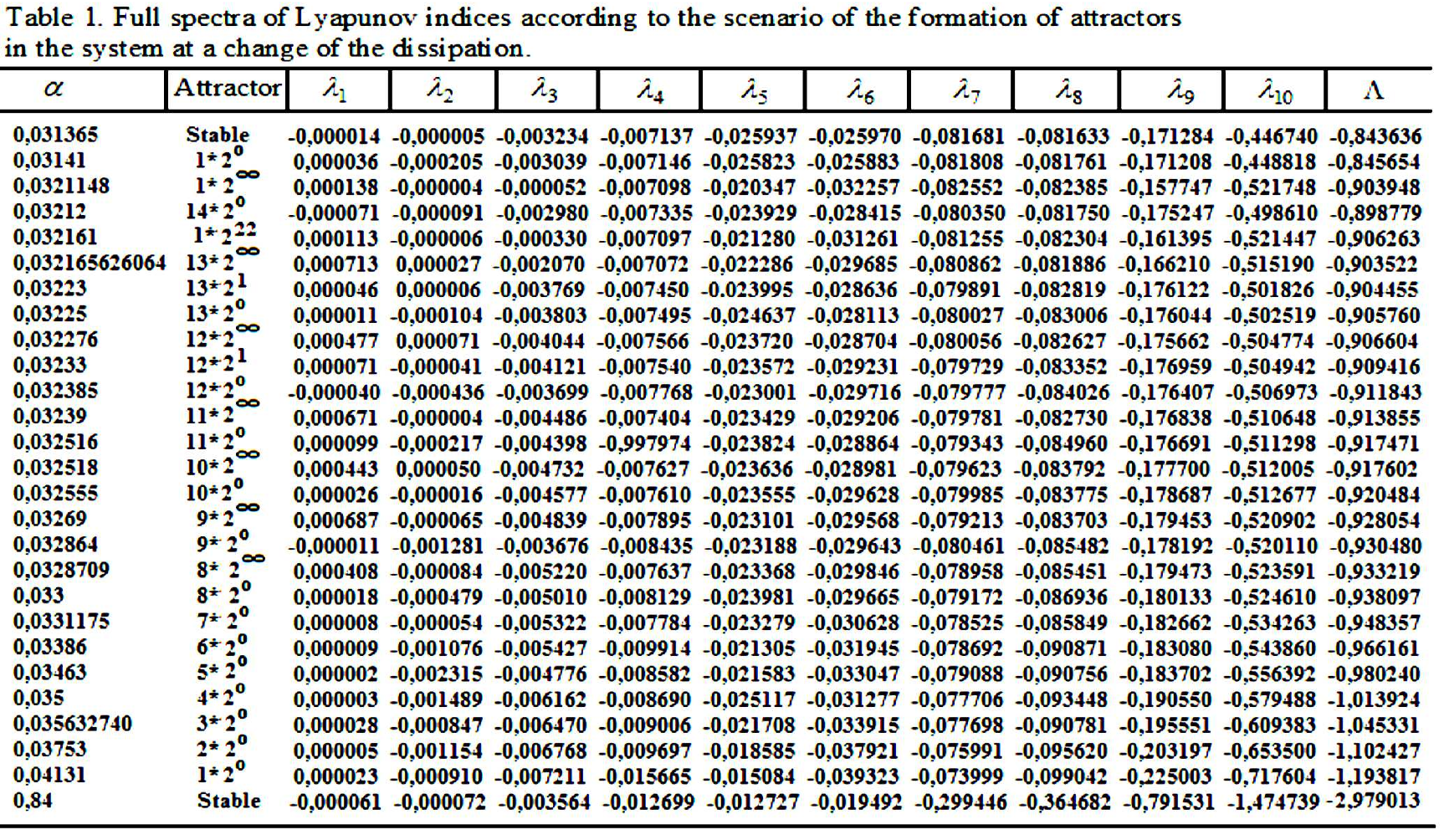}\vskip-3mm
\end{figure*}

\section{Results of Studies}

In work [18], the diagram of states of the system in the parametric
space of input flows of the substrate and oxygen was constructed. By
varying the input flow, it was established that the scenario of the
formation of autoperiodic and chaotic modes is regularly repeated.
Numerical calculations showed that the same scenario is preserved
for fixed flows, but under a change of the dissipation of a kinetic
membrane potential. Such a scenario is presented in Table 1 together
with the spectra of Lyapunov indices for the given modes. As the
dissipation coefficient decreases from 0.84 down to 0.04131, the
stationary state is destroyed, the attractor of a single-valued
autoperiodic mode is formed as a result of the Andronov--Hopf
bifurcation, and then the bifurcations with the doubling of the
period from a single cycle to an 8-fold cycle appear. A subsequent
decrease in the dissipation coefficient leads to the formation of
strange attractors with the corresponding multiplicity between
regular attractors. After the attainment of the 14-fold period, a
single cycle is formed again, and then it passes to the stationary
state. The examples of the corresponding attractors and the kinetic
curves are given in Fig. 1,{\it a--d}. The shape of all regular
attractors $n*2^0$  is analogous to those in Fig. 1,b, where $n = 1,
2,...,14$. Strange attractors $n*2^\infty$, where $n = 8, 9,...,13$,
are analogous to those in Fig. 1,{\it c}. It is of interest that,
after the appearance of a strange attractor $13*2^\infty$ at $\alpha
= 0.032160$, a regular attractor $22*2^\infty$ is formed (see Fig.
1,{\it d}), after which the attractor $14*2^0$ appears further. One
more specific feature of the dynamics of biosystems is shown in Fig.
1,{\it a}. It is revealed at small input flows of the substrate and
oxygen. We observe the appearance of a strange attractor which
differs from the previous ones and possesses a complicated
structure. It arises regularly at various input flows on the
boundary of the transition from a stationary state to $1*2^0$. At
small input flows, in addition to oscillations due to the
desynchronization of the processes of transformation and
accumulation of substrates in the biosystem, there appears the
desynchronization between the processes of respiration and
transformation of the substrate. Two unstable points appear. The
trajectories rotate chaotically around them, by passing from one
center of rotation to another one. By comparing Fig. 1,{\it a} and
Fig. 1,{\it c}, it is worth noting that the chaotic mode in this
biosystem is formed by two means: in the first case, the attractor
creates folds inside itself, whereas a funnel is formed in the
second case. Due to this circumstance, the chaotic motion mixes the
trajectories in the phase space.\looseness=1

In addition to the phase portraits, the figures show the kinetics of one of the
variables of the system. It is seen that the curves differ from one another in
different modes. For strange attractors, the plots represent irregular
oscillations. We indicate a combination of oscillations and jumps. The figures
demonstrate also the dependence of the chaotic kinetics on the initial
conditions.

An important role in the analysis of
the scenario of the formation of various modes is played by Lyapunov indices. For characteristic modes,
Table 1 presents their full spectra $\lambda_1$, $\lambda_2$,...,$\lambda_{10}$,
and the value of their sum $\Lambda = \sum_{j=1}^{10} {\lambda_j}$.
Figure 2,a-d gives the plots of the dependence of $\lambda_1$, $\lambda_2$,
$\lambda_3$, and $\Lambda$  on the dissipation
coefficient $\alpha$ in the interval from 0.0321 to 0.033.

By analyzing the results obtained, we note that all autoperiodic
modes corresponding to regular attractors $n*2^\infty$  have higher
Lyapunov indices practically equal to zero. But the chaotic modes
corresponding to strange attractors $n*2^\infty$ have higher
Lyapunov indices which are positive and greater by one order. It is
seen in Fig. 2,{\it a} how the ``windows of periodicity'' are formed
at $\lambda_1<0$. At the given $\alpha$, the regular attractors
appear, whereas the strange attractors arise outside them. The most
pronounced chaotic modes correspond to maximal peaks of $\lambda_1$.
By the given plot, it is possible to choose beforehand the
corresponding mode of functioning for a bioreactor making no
calculations again.

\begin{figure*}
\includegraphics[width=13cm]{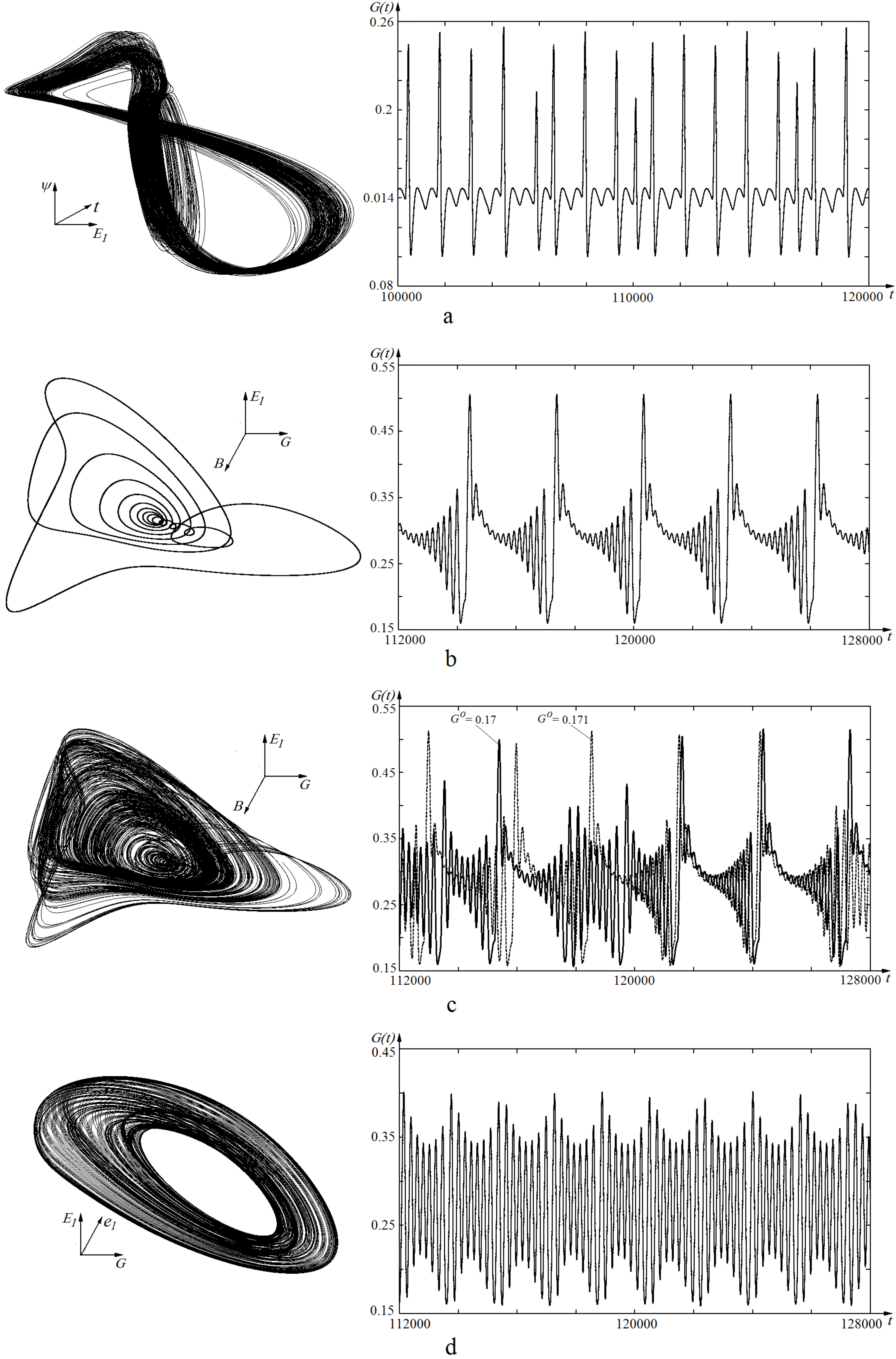}
\caption{Phase portraits and kinetic curves of attractors of the
biosystem: {\it a} -- strange attractor at $\alpha = 0.033$; $G_0 =
0.009$; $O_{20} = 0.00209$; {\it b} -- regular attractor $14*2^0$ at
$\alpha = 0.0321149$; $G_0 = 0.019$; $O_{20} = 0.015$; {\it c} --
strange attractor $13*2^\infty$ at $\alpha = 0.032165626064$; $G_0 =
0.019$; $O_{20} = 0.015$; {\it d} -- regular attractor $22*2^0$ at =
0.032160; $G_0 = 0.019$; $O_{20} = 0.015$ }
\end{figure*}

\begin{figure*}
\includegraphics[width=13cm]{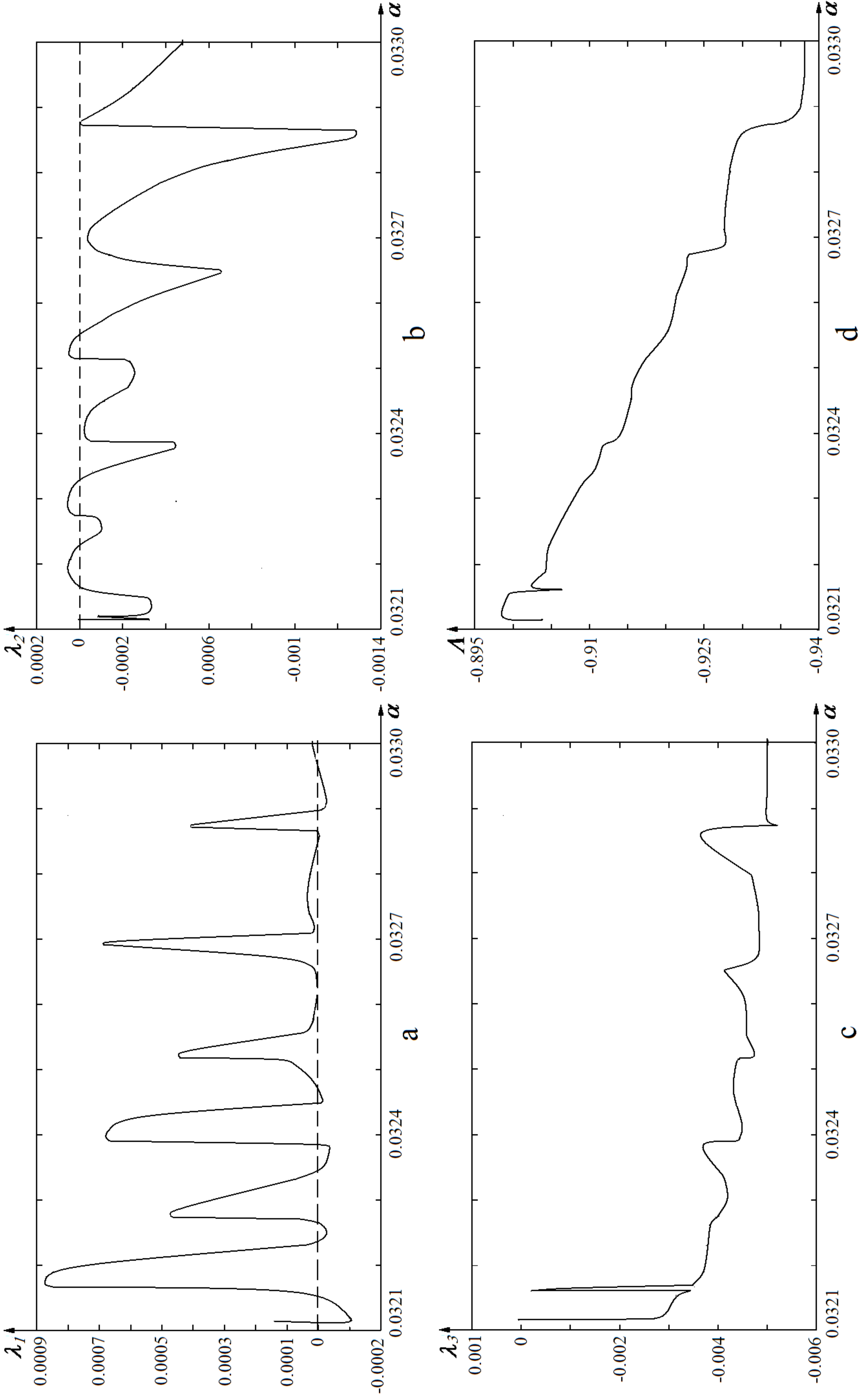}
\caption{Lyapunov indices versus the dissipation coefficient
$\alpha$: {\it a} -- $\lambda_1(\alpha)$; {\it b} --
$\lambda_2(\alpha)$; {\it c} -- $\lambda_3(\alpha)$; {\it d} --
$\Lambda(\alpha)$  }
\end{figure*}

In Fig. 2,{\it b}, we show the plot for the second Lyapunov index.
There, $\lambda_2$ changes accordingly to $\lambda_1$. For strange
attractors, $n*2^\infty$, $\lambda_1>0$, whereas $\lambda_2\approx
0$. That is, the diverging trajectories hold themselves closely to
the given limiting cycle, which preserves the multiplicity of a
strange attractor. The transition from the limiting cycle to the
chaotic mode occurs by means of the intermittence. The kinetics of
the variable in Fig. 2,{\it b}, shows how the periodic limiting cycle is
suddenly broken by the chaotic motion; but then the periodicity is
restored again. On the phase portrait, we can separate a clearly
pronounced region, whose shape is close to that of the disappeared
limiting cycle, relative to which a chaotic trajectory was formed.

The variation of the third Lyapunov index is shown in Fig. 2,{\it
с}. The behavior of $\lambda_3$  is characteristic by that this
index is changed oppositely to $\lambda_1$ and $\lambda_2$. When
these two indices grow, $\lambda_3$  decreases, and {\it vice versa}. For
all $\alpha$, the index $\lambda_3$  is negative.

We may imagine that a 10-dimensional parallelepiped $\overline P^{10}$
is constructed in the given phase space $\overline R^{10}$ at the beginning
of a trajectory on the base of the perturbation vectors
$\overline {b^0_1}$, $\overline {b^0_2}$,... $ \overline {b^0_{10}}$,
which are orthogonal to one another and are normed by one. The value of each
index $\lambda_j$  characterizes a deformation of this parallelepiped along
the corresponding perturbation vector
$\overline {b^i_1}$, $ \overline {b^i_2}$,... $ \overline {b^i_{10}}$,
after $i$ steps of the motion along the trajectory. The parallelepiped spreads along the given vector for a
positive Lyapunov index and shrinks for a negative Lyapunov index.

The sum of all indices $\Lambda$ as a function of $\alpha$ is
given in Fig. 2,{\it d} and in Table 1. This quantity determines
the flow divergence and, hence, the evolution of a phase volume
along a trajectory. For the given dissipative system, the
divergence is negative and increases with the growth of the
dissipation $\alpha$. This means that the phase volume element
shrinks, on the whole, along a trajectory for all values of
$\alpha$. The greater the dissipation, the greater the
contraction. For the regular attractors $n*2^0$, the contraction
is greater than that for the corresponding strange attractor
$n*2^\infty$.

Since we deal with a dissipative system, whose divergence must be
negative in any modes, the phase volume must always shrink. But
$\overline b^i_1 > 0$ in the modes of a strange attractor, and there
occurs the exponential spreading in this direction. Because two
adjacent orbits cannot permanently exponentially diverge, the
strange attractor organizes itself so that it creates a fold (Fig.
1,{\it a}) or a funnel (Fig. 1,{\it c}) in itself, where the mixing
of trajectories is realized. Even a slight deviation of the initial
data influences essentially the evolution of the trajectory, namely
the deterministic chaos is created. Such a chaos characterizes the
appearance of a random nonpredictable behavior of a system
controlled by deterministic laws. We note that, in real biosystems,
the fluctuations are permanently present and, in unstable modes,
create chaotic states. Thus, the given mathematical model adequately
describes stable autoperiodic modes, as well as unstable chaotic
ones.\looseness=1

\section{Conclusions}

Due to the successful development of an algorithm of calculations of the full
spectrum of Lyapunov indices on an ordinary personal computer for a
multidimensional phase space not bounded by the number of variables, we manage
to reliably calculate these indices. This allows one to extend the
possibilities to forecast the dynamics of complicated systems. By the example
of a mathematical model of biosystems, we have found two different scenarios of
the formation of the modes of a strange attractor: the creation of a fold or a
funnel, where the formation of a deterministic chaos is realized. The
self-organization of the phase flow of a strange attractor occurs under the
action of two mutually competitive processes: the exponential extension (of one of
the components, in the given case) and the dissipative contraction of the whole
phase space. Any fluctuation which has appeared there causes the
nonpredictability of the evolution of the system on the whole.




\begin{thebibliography}{99}


\bibitem{1}
L.D. Landau and E.M. Lifshitz, {\it Fluid Mechanics} (Pergamon Press, Oxford, 1975).

\bibitem{2}
A.M. Turing, Phil. Trans. Roy. Soc. Lond. B {\bf 237}, 37 (1952).

\bibitem{3}
G. Nicolis and I. Prigogine, {\it Self-Organization in Nonequilibrium Systems.
From Dissipative Structures to Order through Fluctuations} (Wiley, New York, 1977).

\bibitem{4}
Yu.M. Romanovskii, N.V. Stepanova, and D.S.~Chernavskii, {\it
Mathematical Biophysics} (Nauka, Moscow, 1984) (in Rissian).

\bibitem{5}
T.S. Akhromeyeva, S.P. Kurdyumov, G.G. Malinetskii, and A.A.
Samarskii, Physics Reports {\bf 176}, 189 (1989).

\bibitem{6}
T. Reichenbach, M. Mobilia, and E. Frey, Physical Reviev E {\bf 74},
051907 (2006).

\bibitem{7}
J. W.Pr\"o\ss, R. Schnaubelt, and R. Zacher, {\it Mathematische
Modelle in der Biologie} (Basel, Birkh\"auser,
2008).

\bibitem{8}
V.P. Gachok, V.I. Grytsay, A.Yu. Arinbasarova, A.G.~Me\-dentsev,
K.A. Koshcheyenko, and V.K. Akimenko, Biotechnology and
Bioengineering {\bf 33}, 661 (1989).

\bibitem{9}
V.P. Gachok, V.I. Grytsay, A.Yu. Arinbasarova, A.G.~Me\-dentsev,
K.A. Koshcheyenko, and V.K. Akimenko, Biotechnology and
Bioengineering {\bf 33}, 668 (1989).

\bibitem{10}
V.I. Grytsay, Dopov.  NAN Ukr. No 2, 175 (2000).

\bibitem{11}
V.I. Grytsay, Dopov.  NAN Ukr. No 3, 201 (2000).

\bibitem{12}
V.I. Grytsay, Dopov.  NAN Ukr. No 11, 112 (2000).\vskip1.5mm

\bibitem{13}
V.I. Grytsay, Ukr. Fiz. Zh. {\bf 46}, 124 (2001).\vskip1.5mm

\bibitem{14}
V.V. Andreev and V.I. Grytsay, Matem. Modelir. {\bf 17},  57
(2005).\vskip1.5mm

\bibitem{15}
V.V. Andreev and V.I. Grytsay, Matem. Modelir. {\bf 17}, 3
(2005).\vskip1.5mm

\bibitem{16}
V.I. Grytsay and V.V. Andreev, Matem. Modelir. {\bf 18}, 88
(2006).\vskip1.5mm

\bibitem{17}
V.I. Grytsay, Biofiz. Visn. No 2, 25 (2008).\vskip1.5mm

\bibitem{18}
V.I. Grytsay, Biofiz. Visn. No 2, 77 (2009).\vskip1.5mm

\bibitem{19}
A.A. Akhrem and Yu.A. Titov, {\it Steroids and Microorganisms}
(Nauka, Moscow, 1970) (in Russian).\vskip1.5mm

\bibitem{20}
A.G. Dorofeev, M.V. Glagolev, T.F. Bondarenko, and N.S. Panikov,
Mikrobiol. {\bf 61}, 33 (1992).\vskip1.5mm

\bibitem{21}
I. Shimada and T. Nagashima, Progr. Theor. Phys. {\bf 61}, 1605
(1979).\vskip1.5mm

\bibitem{22}
M. Holodniok, A. Klic, M. Kubicek, M. Marek, Metody Anal\'{y}zy
Nelinearnich Dynamickych Modelu (Academia, Praha, 1986).\vskip1.5mm

\bibitem{23}
S.P. Kuznetsov, {\it Dynamical Chaos} (Nauka, Moscow, 2001) (in
Russian).\vskip1.5mm

\begin{flushright}
{\footnotesize Received 30.09.09}
\end{flushright}
\end{thebibliography}
\end{document}